\documentclass[aps,prl,superscriptaddress,twocolumn,longbibliography,floatfix]{revtex4-2}
\usepackage{units}
\usepackage{amsmath}
\usepackage{amsthm}
\usepackage{amssymb}
\usepackage{graphicx}
\usepackage[english]{babel}
\usepackage[utf8]{inputenc}
\usepackage{mathtools}
\usepackage{color}
\usepackage{xcolor}
\usepackage{bbold}
\usepackage{braket}
\usepackage{tabularray}
\usepackage{subfigure}
\usepackage[version=3]{mhchem}
\usepackage{dsfont}

\usepackage{multirow}

\definecolor{myurlcolor}{rgb}{0,0,0.7}
\definecolor{myrefcolor}{rgb}{0.8,0,0}

\usepackage[
	breaklinks,
	pdftex,
	colorlinks=true, 
	linkcolor=myrefcolor,
	citecolor=myurlcolor,
	urlcolor=myurlcolor
]{hyperref}
\usepackage{cleveref}

\usepackage[linesnumbered, ruled,vlined]{algorithm2e}

\usepackage{ulem}
\makeatletter
\def\uwave{\bgroup \markoverwith{\lower3.5\p@\hbox{\sixly \textcolor{red}{\char58}}}\ULon}
\font\sixly=lasy6 
\makeatother

\graphicspath{{./images/},{./imagesAppendix/},{./Plots/}}

\renewcommand{\eqref}[1]{Eq.~(\ref{#1})} 

\def\app#1#2{%
  \mathrel{%
    \setbox0=\hbox{$#1\sim$}%
    \setbox2=\hbox{%
      \rlap{\hbox{$#1\propto$}}%
      \lower1.1\ht0\box0%
    }%
    \raise0.25\ht2\box2%
  }%
}

\ifx\proof\undefined
\newenvironment{proof}[1][\protect\proofname]{\par
	\normalfont\topsep6\p@\@plus6\p@\relax
	\trivlist
	\itemindent\parindent
	\item[\hskip\labelsep\scshape #1]\ignorespaces
}{%
	\endtrivlist\@endpefalse
}
\providecommand{\proofname}{Proof}
\fi
\makeatother
\newcommand{\abs}[1]{\left|#1\right|}
\newcommand{\idg}[1]{{\bfseries #1)}}

\providecommand{\ftname}{ft}
\providecommand{\theoremname}{Theorem}
\providecommand{\claimname}{Claim}
\providecommand{\lemmaname}{Lemma}
\providecommand{\definitionname}{Definition}

\definecolor{KB}{rgb}{0.4,0.3,0.9}

\definecolor{THc}{rgb}{0.9,0.3,0.2}

\newcommand{\gb}[1]{{\color{black} #1}}

\newcommand{\la}[1]{{\color{black} #1}}

\newcommand{\pk}[1]{{\color{black} #1}}

\newcommand{\blue}[1]{{\color{black} #1}}

\newcommand{\sectionMain}[1]{
\let\oldaddcontentsline\addcontentsline
\renewcommand{\addcontentsline}[3]{}
\section{#1}
\let\addcontentsline\oldaddcontentsline
}

\allowdisplaybreaks

\begin{document}

\title{\blue{Rydberg atoms for electric field gradiometry}
}

\author{Philip Kitson}
\affiliation{Quantum Research Center, Technology Innovation Institute, P.O. Box 9639 Abu Dhabi, UAE}
\affiliation{Dipartimento di Fisica e Astronomia ``Ettore Majorana'', Universit\'a di Catania, Via S. Sofia 64, 95123 Catania, Italy}
\affiliation{INFN-Sezione di Catania, Via S. Sofia 64, 95123 Catania, Italy}
\author{Wayne J. Chetcuti}
\affiliation{Quantum Research Center, Technology Innovation Institute, P.O. Box 9639 Abu Dhabi, UAE}
\affiliation{Université Grenoble Alpes, CNRS, LPMMC, 38000 Grenoble, France}
\author{Gerhard Birkl}\affiliation{Technische Universit\"at Darmstadt, Institut f\"ur Angewandte Physik, Schlossgartenstra\ss e 7, 64289 Darmstadt, Germany}
\affiliation{Helmholtz Forschungsakademie Hessen f\"ur FAIR (HFHF), GSI Helmholtzzentrum für Schwerionenforschung, 64291 Darmstadt}
\author{Luigi Amico}
\affiliation{Quantum Research Center, Technology Innovation Institute, P.O. Box 9639 Abu Dhabi, UAE}
\affiliation{Dipartimento di Fisica e Astronomia ``Ettore Majorana'', Universit\'a di Catania, Via S. Sofia 64, 95123 Catania, Italy}
\affiliation{INFN-Sezione di Catania, Via S. Sofia 64, 95123 Catania, Italy}
\author{Juan Polo}
\affiliation{Quantum Research Center, Technology Innovation Institute, P.O. Box 9639 Abu Dhabi, UAE}

\begin{abstract}
    \blue{We propose a quantum sensor for electric fields based on networks of Rydberg atoms. The sensing mechanism exploits the strong dependence of the Rydberg blockade on the applied electric field near a Förster resonance. In this regime, variations of the electric field across the array lead to local changes in the blockade radius. Therefore, owing to its spatially distributed architecture, the device can operate as a gradiometer. Our analysis shows that our scheme enables detection of spatial variations in the electric field with a resolution of few $\mu$m. We analyse the dynamics of Rydberg excitations for systems with different spatial geometries and electric field configurations to establish the relation between the applied field and the blockade response. For spatially inhomogeneous fields, we also provide another observable, density–density correlations, that can probe the field's spatial structure.}
\end{abstract}

\maketitle





\textit{Introduction --} Rydberg atoms, characterised by electrons occupying states with a high principal quantum number $n$, are known to possess unique properties, such as an increased orbital radius (scaling as $n^{2}$), heightened polarisability ($n^{7}$), and strong dipole-dipole interactions~\cite{wu2021a, adams2019rydberg, gallagher1988rydberg}. The dipolar interaction can be effectively scaled as either $R^{-3}$ or $R^{-6}$, depending on the coupled states involved, i.e., states with either opposite or same parity, respectively. These strong interactions can cause shifts in Rydberg energy levels, leading to an excitation blockade effect in nearby atoms when excited with narrow-band radiation~\cite{jaksch2000fast, lukin2001dipole}. One way to manipulate the atom-atom interactions is by applying an electric field, which results in a differential Stark shift of the levels. 
Because of the latter, the  transitions between {\it pairs} of Rydberg states, each in different atoms, can become resonant.
In this case, an increased strength of resonant dipole-dipole interactions turns out to occur that lead to an enhanced blockade radius. In  analogy with the energy transfer  originally discussed for organic molecules~\cite{forster1948zwischenmolekulare,sahoo2011forster},  such a regime is known as a Förster resonance~\cite{anderson2002dephasing, walker2005zeros, vogt2007electric, vanDitzhuijzen2008spatially, ryabtsev2010observation, nipper2012highly, nipper2012atomic}.
%
These resonances have been mostly  investigated for two atoms (two-body Förster resonances). 
Efforts have been used to explore three-body Förster resonances~\cite{faoro2015borromean, tretyakov2017observation, ryabtsev2018coherence, cheinet2020three, ryabtsev2025investigation}
with applications for three-qubit quantum gates~\cite{beterov2018fast} and measurements of the dipole-dipole interaction~\cite{ryabtsev2025splitting}. Local addressability is also possible using inhomogeneous light fields such as focused laser beams~\cite{emperauger2025benchmarking}.

\begin{figure}
    \centering
    \includegraphics[width=\linewidth]{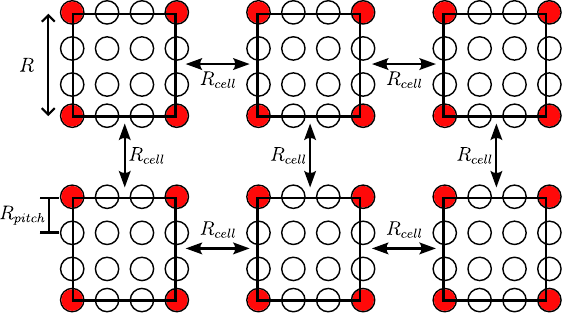}
    \caption{\color{black} 
    \textit{Schematic for an electric field gradiometer.} Operational principle of the Rydberg-atom electric field sensor based on a 2D optical-tweezer array with consisting of square cells (black lines) of four Rydberg atoms (red dots) with nearest-neighbour tweezer spacing $R$. The separation between cells, $R_\text{cell}$, is chosen much larger than $R$ to suppress inter-cell interactions, so that each cell acts as an independent sensing unit probing the local electric field $E$. In practice, the external electric field modifies the Rydberg blockade radius (near the Förster resonance) for a fixed interatomic distance $R$, producing characteristic changes in the population of the many-body excitations. The local electric field can thus be inferred from measurements of the Rydberg excitation probabilities, with spatial resolution given by $R_\text{cell}$ (which can be reduced to $R_{pitch}$ - the separation of adjacent optical tweezers - using the specific protocol described in the main text).
    }
    \label{fig.tweezer_arrangement}
\end{figure}

Clearly, the effect of external fields on atomic systems can provide a physical principle to be exploited in advanced quantum sensing applications~\cite{degen2017quantum, adams2019rydberg}. 
For electric field sensors, Rydberg atoms offer an excellent platform due to their notable polarizability~\cite{degen2017quantum, zhang2024rydberg, yuan2023quantum}. The platform has demonstrated sensitivities of several $\mu$V/cm/$\sqrt{\text{Hz}}$~\cite{facon2016singlerydberg,kumar2017rydberg} and can detect fields 
up to the THz range~\cite{wade2016real, meyer2020assessment}. 
For scalability and control over interatomic interactions, one can arrange atoms in arrays of optical tweezers generating one-, two-, and three-dimensional (1D, 2D, and 3D) networks \cite{pause2024supercharged, schlosser2023talbot} of quantum sensors with selectable separations down to the single-micron range~(Fig.~\ref{fig.tweezer_arrangement}). Recently, a 2D network of individual rubidium atoms has been used as a sensor for a spatially varying magnetic field~\cite{schaeffner2024quantum}. 

\begin{figure*}
    \centering
    \includegraphics[width=\linewidth]{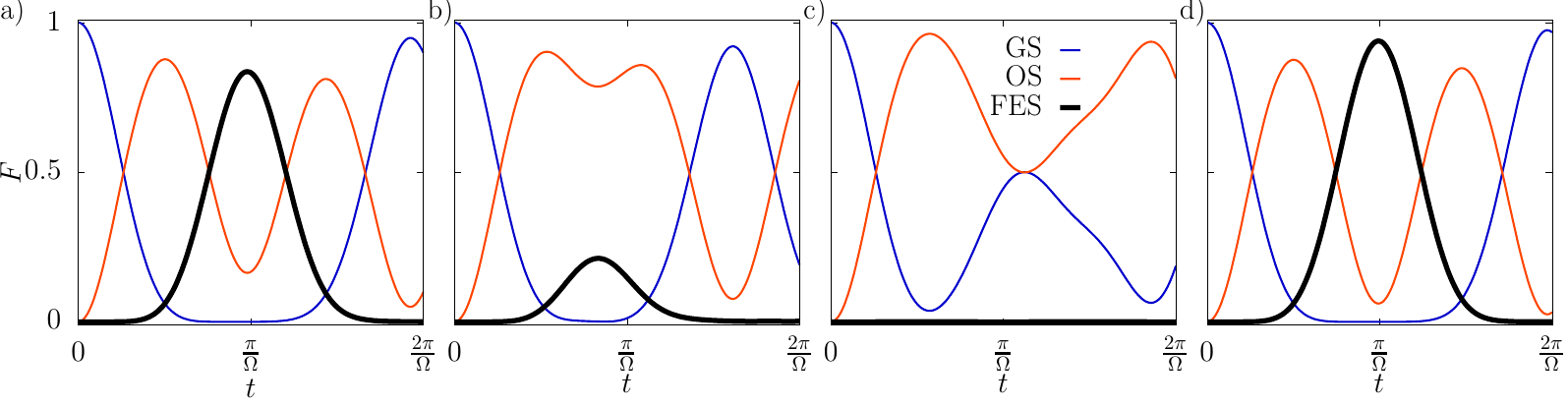}
    \caption{The fidelity $F$ of the basis states for four interacting atoms in a square-cell configuration having a separation $R$ = 15 $\mu$m over a Rabi period, $t \in (0, \tau = 2\pi/\Omega)$ for different electric fields $E$ with $\Omega$ corresponding to the Rabi frequency. GS \gb{(blue)} represents the ground state; OS \gb{(red)} is the sum of the states with one, two, and three excitations; and FES \gb{(black)} is the fully excited state. \idg{a} electric field, $E = 0$ with $F_{max} = 0.836$ being the maximum projection over the fully excited state (see~\eqref{eq.f_max}). \idg{b} $E = 25$ mV/cm with $F_{max} = 0.212$. \idg{c} $E = E_{res}= 29.787$ mV/cm (corresponding to the resonant electric field) with $F_{max} = 0.001$. \idg{d} $E = 50$ mV/cm with $F_{max} = 0.939$.}
    \label{fig.full_state_dynamics_N_3}
\end{figure*}


{\color{black}
In this work, we present a complementary approach to existing gas-cell Rydberg atom sensors, which are primarily sensitive to spatially uniform electric fields~\cite{facon2016singlerydberg,kumar2017rydberg}. Our scheme exploits Rydberg states for which an external electric field increases the blockade radius when operating near a Förster resonance. As a consequence, spatially varying electric fields can be detected on micrometer length scales. To illustrate the sensor’s functionality, we consider a system with fixed interatomic separation while varying the applied electric field, which modifies the blockade radius. The resulting field-induced change in the blockade radius produces characteristic many-body dynamics that can be probed through measurements of the fully excited Rydberg-state population and density–density correlators.

Based on these concepts, we propose a specific design for a quantum electric-field sensor using a typical 2D network of Rydberg atoms, as illustrated in Fig.~\ref{fig.tweezer_arrangement}. The core tweezer array has equal vertical and horizontal spacing $R_{pitch}$, with atoms occupying selected sites to form a square pattern. The number of empty tweezers between neighbouring cells is such that interactions between atoms in different cells are suppressed. Through this, we show how multiple atom network configurations enable the measurement of the \textit{absolute magnitude} of the electric field, with sensitivity and operating range determined by the F\"orster resonance and their pitch separation. }


\textit{Models and Methods --} Single atoms with primarily one or two electrons in their outer shell can be driven from the ground $\ket{g}$ to a Rydberg state $\ket{\alpha}$ using a resonant field that couples these two states. Assuming a coupling strength characterised by the Rabi frequency $\Omega$, the system undergoes Rabi oscillations between the two states~\cite{dunning2016recent, adams2019rydberg}. When multiple atoms are driven simultaneously, long-range interactions emerge due to the large dipole moment of Rydberg atoms. For $N$ atoms, such a system can be modelled via the Hamiltonian~\cite{browaeys2020many, adams2019rydberg}

\begin{equation}
    H = \frac{\Omega}{2}\sum_{i}^{N} \, \sigma_{i}^{x} - \Delta\sum_{i}^{N} \, n_{i} + \sum_{i \neq j}^{N} V_{i, j} \, n_{i}n_{j}.
    \label{eq:two_level_hamiltonian}
\end{equation}

\noindent The Pauli matrices are denoted by $\sigma_{i}^{u}$ (with $u = x, y, z$), and the parameters are $\Delta$ for the laser detuning and $V_{i, j}$ for the interaction strength between Rydberg atoms $i$ and $j$. The local number operator defined as $n_{i} = \frac{1}{2}(\sigma_{i}^{z} + \mathds{1})$ counts atoms in the Rydberg state. Due to the interaction-induced energy level shift, nearby atoms are prevented from being excited to the Rydberg state -- an effect known as the Rydberg blockade
~\cite{lukin2001dipole, jaksch2000fast, urban2009observation, gatan2009observation}. 
Here we set $\Delta=0$ to explore the effects of the blockade.

The Hamiltonian in~\eqref{eq:two_level_hamiltonian} describes the physics when an atom is excited to a specific Rydberg state $\ket{\alpha}$. We now extend this description to the case where additional nearby Rydberg states are energetically accessible. Consider two identical atoms separated by a distance $R \equiv R_{1,2}$, each with available Rydberg states $|\alpha\rangle$, $|\alpha'\rangle$, and $|\alpha''\rangle$. Whenever the pair state $|\alpha,\alpha\rangle \equiv |\alpha\rangle_1 \otimes |\alpha\rangle_2$ is nearly degenerate with $|\alpha',\alpha''\rangle$ and $|\alpha'',\alpha'\rangle$, it can couple via the dipole–dipole interaction. In fact, due to the exchange symmetry of the dipole–dipole interaction~\cite{nipper2012atomic}, $|\alpha,\alpha\rangle$ couples only to the symmetric superposition $|+\rangle = (|\alpha',\alpha''\rangle + |\alpha'',\alpha'\rangle)/\sqrt{2}$. Therefore, in the subspace $\{|\alpha,\alpha\rangle, |+\rangle\}$, the interaction can be described by an effective two-level Förster Hamiltonian:

\begin{equation}
    V^{(\alpha)} = \begin{pmatrix}
    \delta & C_{3}/R^{3} \\
    C_{3}/R^{3} & 0 
\end{pmatrix},
\label{eq:pair_state_hamiltonian}
\end{equation}


\noindent where we define the so-called energy defect $\delta = 2\epsilon_{\alpha} - \epsilon_{\alpha^{\prime}} - \epsilon_{\alpha^{\prime\prime}}$, with $\epsilon_{\eta}$ denoting the single-atom energy of state $\eta$. The term $C_{3}/R^{3}$ represents the resonant dipole–dipole matrix element between the pair states, with $C_{3}$ being dependent on the dipole matrix elements of the pair states.
%
%
The  effective interaction between two atoms is provided by the lowest energy eigenvalue of~\eqref{eq:pair_state_hamiltonian}:

\begin{equation}
    V = \frac{\delta - \sqrt{\delta^{2} + 4C_{3}^{2}/R^{6}}}{2},
    \label{eq:eigenvalues}
\end{equation}

\noindent whereby, the competition between $\delta$ and $C_{3}/R^{3}$ determines the nature of the interaction term. In the case  $\delta \gg C_{3}/R^{3}$,~\eqref{eq:eigenvalues} can be simplified to $V \simeq -C_{3}^{2}/\delta R^{6}$, which gives rise to Van der Waals-like interactions (VdW). Conversely, if $\delta \ll C_{3}/R^{3}$ then $V \simeq C_{3}/R^{3}$, resulting in dipole-dipole interactions. When $\delta \approx C_{3}/R^{3}$, the system undergoes a crossover between the two types of interactions (see Appendix).

In the presence of an electric field $E$, the energy levels experience a Stark shift that alters both $\delta(E)$ and $C_{3}(E)$. Therefore, by gradually increasing the electric field towards the F\"orster resonance, the nature of interaction changes from VdW to dipole-dipole, \textit{resulting in an enlarged  blockade radius}~\cite{reinhard2008effect}. 
Here, we consider states that have been previously shown to display a Förster resonance in $^{87}$Rb, e.g., $\ket{\alpha} = \ket{59D_{3/2}}$, $\ket{\alpha^{\prime}} = \ket{61P_{1/2}}$, and $\ket{\alpha^{\prime\prime}} = \ket{57F_{5/2}}$ at the electric field $E \approx 32~$mV/cm with a magnetic field $B_{z} = 3 \, $G~\cite{ravets2014coherent}. For these states, we utilise the Alkali Rydberg Calculator (ARC)~\cite{sibali2017ARC} and obtain $C_{3}(E)$ and $\delta(E)$ at zero magnetic field $B_{z} = 0$ G. In doing so, we show the relative energies of 
$\ket{59D_{3/2},59D_{3/2}}$ and $\ket{57F_{5/2},61P_{1/2}}$
in \gb{Fig. \ref{fig:Energy_Levels_in_E_Field}} of the Appendix.

The time dynamics can be obtained through the Liouville–von Neumann equation:

\begin{equation}
    \partial_{t}\rho(t) = -i \, \left[H, \, \rho(t)\right],
    \label{eq.von_neumann}
\end{equation}

\noindent where the density matrix is defined as $\rho(t) = \ket{\Psi(t)}\bra{\Psi(t)}$ with $\ket{\Psi(t)}$ corresponding to the state of the system. The 
influence
of the electric field is accounted for through the effective interactions \eqref{eq:eigenvalues} in \eqref{eq:two_level_hamiltonian}.


\textit{Dependence of the Rydberg blockade on the electric field --} 
First, we show how the blockade radius changes under external homogenous electric fields in arrays of Rydberg atoms. We consider an experimentally realistic 2D tweezer network with a fundamental pitch of $R_{pitch} = 5 \, \mu$m 
{\color{black}
in which $N = 4$ rubidium atoms are placed in a cell configuration - as shown in Fig.~\ref{fig.tweezer_arrangement} - at a distance $R = 15 \, \mu$m and subjected to a uniform electric field.} The state dynamics in the Fock basis, represented by $\ket{n_{1}, \hdots, n_{N}}$ with $n_i \in {0,1}$, is shown in Fig.~\ref{fig.full_state_dynamics_N_3}
by plotting the fidelity $F=|\langle n_{1}, \hdots, n_{N}|\Psi(t)\rangle |^{2}$ between  $\ket{\Psi(t)}$ and $\ket{n_{1}, \hdots, n_{N}}$. For electric fields ($E = 0$ mV/cm (Fig.~\ref{fig.full_state_dynamics_N_3}(\textbf{a})) and $E = 50$ mV/cm (Fig.~\ref{fig.full_state_dynamics_N_3}(\textbf{d})) far away from the Förster resonance, the system's population oscillates between the ground and fully excited state $\ket{1_{1}, \hdots, 1_{N}}$, as the resulting blockade radius is smaller than the atom separation. However, for electric fields (e.g., $E = 25$ mV/cm (Fig.~\ref{fig.full_state_dynamics_N_3}(\textbf{b}))) approaching the Förster resonance, the population of
$\ket{1_{1}, \hdots, 1_{N}}$
decreases because the energy defect becomes smaller than the coupling strength.
Consequently, the blockade radius increases and becomes comparable to the atom separation. At the Förster resonance ($E = E_{res} = 29.787$ mV/cm with $B_{z} = 0 \,$ G (Fig.~\ref{fig.full_state_dynamics_N_3}(\textbf{c}))), the blockade radius reaches its maximum value and surpasses the atom separation. As a result, the probability of the system being fully excited drops to nearly zero.

To quantitatively describe the reduction in the probability of the fully excited state around the Förster resonance, we 
compute the maximum value of the fidelity between  $\ket{\Psi(t)}$ and $\ket{1_{1}, \hdots, 1_{N}}$ 
over a Rabi period $t \in (0, \tau = 2\pi/\Omega)$ defined as 
\begin{equation}
    F_{max} = \max_{t \in (0, \tau)} \, (|\langle 1_{1}, \hdots, 1_{N}|\Psi(t)\rangle |^{2}).
    \label{eq.f_max}
\end{equation}

\begin{figure}[h!]
    \centering
    \includegraphics[width=\linewidth]{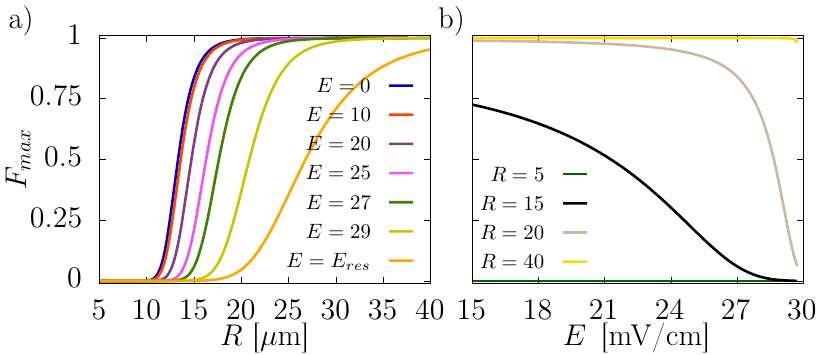}
    \caption{\idg{a} The maximum fidelity for populating the fully excited state, $F_{max}$ (~\eqref{eq.f_max}) for $N = 4$ in a square formation over the time period $t \in (0, \, 2\pi/\Omega)$. The atoms have a fixed separation which increases from $R = 5 \, \mu$m to $R = 40 \, \mu$m for the different electric fields $E = (0, \, 10, \, 20, \, 25, \, 27, \, 29, \, E_{res})$ mV/cm.
    \idg{b} $F_{max}$ as a function of a uniform electric field across all $N = 4$ atoms for fixed separations $R = (5, \, 15, \, 20, \, 40) \, \mu$m.}
    \label{fig_f_max_N_4}
\end{figure}

In Fig.~\ref{fig_f_max_N_4}\textbf{(a)}, we showcase $F_{max}$
for a uniform electric field $E$ across the atoms. At small atom spacings, the value of $F_{max}$ is small as
the blockade radius exceeds the separation. However, as 
the atom distance increases beyond the blockade radius,
$F_{max}$ also rises.
\gb{For different values of $E$, this} increase happens at different atom separations as the blockade is larger for electric fields closer to the Förster resonance. Furthermore, the gradient of $F_{max}$ changes due to the electric field. This is a result of the change in nature of the interactions from $R^{-3}$ to $R^{-6}$, which occurs at short atom-atom separation and more abruptly for $E = 0$ compared to $E = E_{res}$ (see more details in the Appendix).

\begin{figure*}[ht!]
    \centering
    \includegraphics[width=0.85\linewidth]{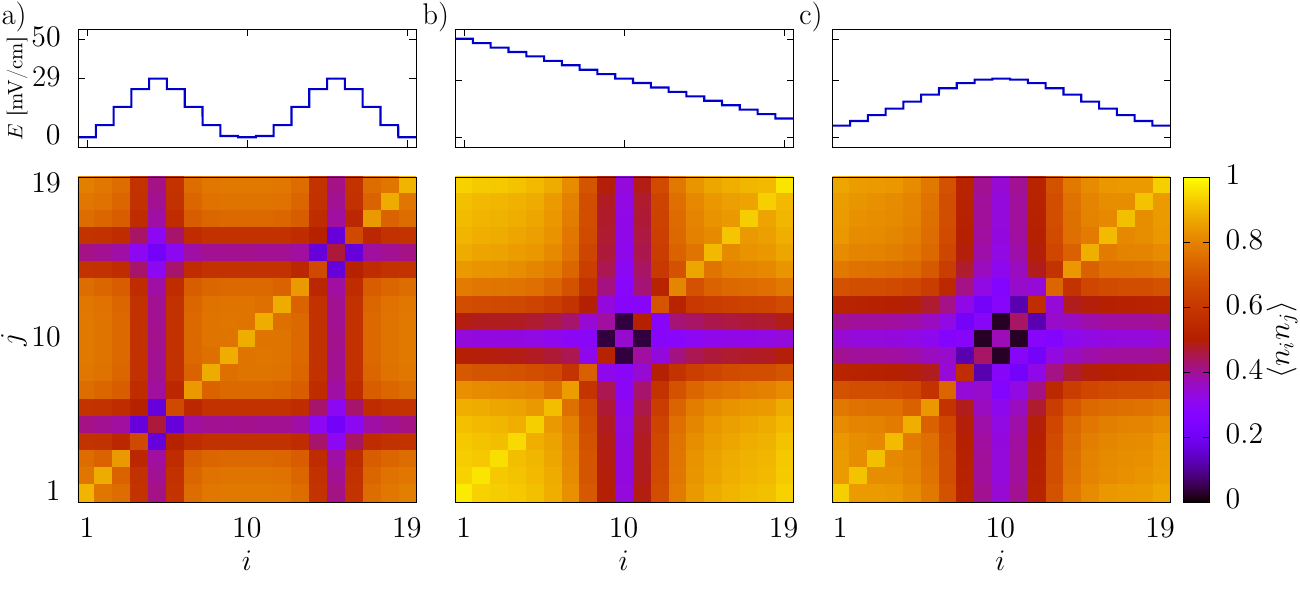}
    \caption{\textbf{Top:} The applied electric field configuration $E/E_{res}$ across $N = 19$ atoms (blue line), which are separated by a distance $R = 15 \, \mu$m. 
    \textbf{Bottom:} The corresponding density-density correlator $\langle n_{i}n_{j}\rangle$ for the above electric field configurations recorded at maximum population of the fully excited state. \idg{a} Sinusoidal function with two resonant peaks, located at $ i, j = 5$ and $i, j = 15$. \idg{b} Gradient field with $E=E_{res}$ for atom $i,j=10$. \idg{c} Gaussian function mimicking the AC Stark shift of a laser beam.}
    \label{fig:density_density_N_19}
\end{figure*} 
Alternatively, 
we can look at $F_{max}$ at fixed atom distances in homogeneous electric fields (Fig.~\ref{fig_f_max_N_4}\textbf{(b)}).
For $R = 5 \, \mu$m, $F_{max}$ remains low for all electric field values as this atom spacing
is much smaller than the blockade radius. Conversely, for $R = 40 \, \mu$m, the separation is much larger than the blockade radius at all electric fields considered; therefore, $F_{max}$ remains high\pk{, providing a suitable lower bound for $R_{cell}$ (see Fig.~\ref{fig.tweezer_arrangement})}. At the intermediate values $R = 15 \, \mu$m and $R = 20 \, \mu$m $F_{max}$ decreases as the electric field approaches the Förster resonance, since the blockade exceeds the atom distance. 
The non-linear behaviour of these curves stems from the way $F_{max}$ is shifted in Fig.~\ref{fig_f_max_N_4}\textbf{(a)}, which result in a steeper gradient at the larger atom spacings $R = 20 \, \mu$m.

\blue{
\begin{table*}[ht!]
\centering
\begin{tabular}{|l|c|c|c|c|}
\hline
 & Sensitivity & Operating Range & Resolution & Repetitions \\
\hline
\multirow{3}{*}{$R = 15\,\mu$m} 
& \multirow{3}{*}{$\approx 5.730\,\text{mV/cm}\times\sqrt{\tau_{cyc}}$} & \multirow{3}{*}{20.36 - 27.25 $\text{mV/cm}$} & $1.5\,\mu$m & 3240 \\
\cline{4-5}
 &  &  & $5\,\mu$m & 960 \\
\cline{4-5}
 &  &  & $40\,\mu$m & 120 \\
\hline
\multirow{3}{*}{$R = 20\,\mu$m} 
& \multirow{3}{*}{$\approx 0.238\,\text{mV/cm}\times\sqrt{\tau_{cyc}}$} & \multirow{3}{*}{28.12 - 29.63 $\text{mV/cm}$} & $1.5\,\mu$m & 3240 \\
\cline{4-5}
 &  &  & $5\,\mu$m & 960 \\
\cline{4-5}
 &  &  & $40\,\mu$m & 120 \\
\hline 
\end{tabular}
\caption{\blue{Sensor performance metrics: sensitivity, operating range, resolution, and repetitions. The sensitivity calculation is provided in the appendix, and the operating range indicates the span between the smallest and largest electric fields the sensor can detect. The resolution varies depending on the separation of the optical tweezers; we consider three values: the first is the state-of-the-art positioning capability of the tweezers~\cite{holman2026trapping}, the second is the standard tweezer separation ($R_\text{pitch}$) used in this work, and the third is the separation of the cells ($R_\text{cell}$). 
The number of repetitions for the experiment is calculated as the product of the number of measurements needed ($n_\mathrm{mes} = 120$ runs - see appendix) and the number of intermediate atom positions required, being determined through \gb{$(R_\text{cell} + R - 1)/R_\text{pitch}$}  from Fig.~\ref{fig.tweezer_arrangement}.
Estimates for typical cycle times $\tau_{cyc}$ are given in \cite{schaeffner2024Bfieldsensor}.
} }
\label{Table_metrics_square}
\end{table*}
}
{\color{black}
While \gb{this configuration ($N = 4$ atom arranged in a square)} captures the main features of the Rydberg blockade’s dependence on the applied electric field and atom separation, alternative geometrical arrangements and/or varying number of atoms can be used to further investigate this dependence. In the Appendix, we examine how both factors influence $F_{max}$, considering a linear configuration and increasing number of atoms from $N= 2$ to $N = 5$.
}

\textit{Density-Density Correlator --} 
{\color{black} The main features of interactions in Rydberg systems are mediated through a two-body process; therefore, using the two-body density-density correlator $\langle n_i n_j \rangle$ offers deeper insight into the particles' interactions. This quantity serves as a direct and experimentally accessible indicator of spatial correlations and provides a clear signature of the Rydberg blockade phenomenon. Our setup depends on the changes of the blockade radius; thus, this measure directly links to the applied electric fields. To illustrate this dependence, we now consider a system of $N = 19$ atoms arranged linearly with an interatomic separation of $R = 15 \, \mu$m, and compute the correlator for various electric field configurations at the moment when the fidelity of the fully excited state ($\ket{1_{1}, \dots, 1_{19}}$) is maximum (Fig.~\ref{fig:density_density_N_19}). In this case, we focus on a one-dimensional array where the electric field varies along the array. Diagonal terms represent the expectation value of the density at each site, which, for atoms subject to electric fields close to resonance, takes smaller values compared to fields far from resonance. Consequently, the local density expectation value provides a signature of the overall shape of the electric field.
}


The super- and sub-diagonal correlator terms (symmetric under $i \leftrightarrow j$) indicate whether neighbouring atoms can be simultaneously excited. 
Small $\langle n_i n_j\rangle$ values signal fields near the Förster resonance, where the blockade radius exceeds the separation and suppresses excitations, while large $\langle n_i n_j\rangle$ values correspond to fields far from resonance, where multiple excitations are allowed. To illustrate this dependence, in Fig.~\ref{fig:density_density_N_19} we consider several electric-field configurations and calculate the resulting correlations. In Fig.~\ref{fig:density_density_N_19}\textbf{(a)}, a spatially sinusoidally varying DC electric field produces correlator minima at resonant positions. A constant negative gradient (Fig.~\ref{fig:density_density_N_19}\textbf{(b)}) yields nearly equal correlators about the resonance, though slight differences arise because $|\delta|$ and $C_{3}$ are not fully symmetric with respect to $\delta=0$. Figure~\ref{fig:density_density_N_19}\textbf{(c)} shows the response to a Gaussian-shaped field, emulating the AC Stark shift from a focused laser beam~\cite{emperauger2025benchmarking}. All these cases show that $\langle n_{i}n_{j}\rangle$ can capture the overall shape of spatially varying electric fields as this measurement directly reflects interaction-induced spatial correlations.

\color{black}


\textit{Sensing of Electric Fields --}
{\color{black}
In this section, we show how the electric field dependence of the Rydberg blockade can be exploited to realize an electric field quantum sensor capable of probing measurements of spatially varying electric fields $E(x,y)$ with resolutions of the order of micrometers.
Using the geometry outlined in Fig.~\ref{fig.tweezer_arrangement}, we consider a two-dimensional tweezer array with pitch $R_{\text{pitch}}$, in which $N = 4$ atoms separated by a distance $R$ form a square sensor cell, while the distance between cells, $R_\text{cell}$, is set to be much larger than the atom-atom separation such that the Rydberg interactions between different cells is negligible. The electric-field-induced change in the blockade radius is quantified by the maximum fidelity $F_{max}$ and can be effectively exploited to extract direct information about the electric field experienced by the atoms within a given cell. In an experimental setting, the fidelity of the state $|1_1,\cdots,1_N\rangle$ can be directly accessed by measuring the Rydberg excitation state of each atom~\cite{urban2009observation,saffman2010quantum}.

In order to estimate the electric field $E(x,y)$ from the probability of the fully excited state, we propose the use of Bayesian statistics. After an initial calibration that provides the relation between the fully excited state probability and the electric field, we can perform multiple measurements of the local population of the Rydberg states and use Bayesian statistics to estimate the local electric field in each cell, with a degree of certainty associated with the number of measurements/repetitions $n_\text{rep}$. This yields $E(x,y)$ with a spatial resolution of $R_\text{cell}$, so each cell provides us with the average electric field over those 4 atoms. \pk{Higher spatial resolution of $R_{pitch}$ can be achieved by simply selecting new cell positions whose displacement with respect to the previous ones is given by $R_{pitch}$.} 
To obtain this increased resolution requires one to perform a larger number of measurements equal to the number of shifts required to cover the full grid.

Metrics for the proposed sensor, assuming typical experimental parameters of $R_\text{pitch} \approx 5\mu$m for $^{87}$Rb atoms, can be found in Table~\ref{Table_metrics_square}. The electric-field sensitivity is obtained from the field uncertainty inferred by error propagation, where the uncertainty in the measured fidelity, extracted from the Bayesian analysis after $n_\mathrm{mes}=120$ measurements, is divided by the local slope $|dF_{max}/dE|$ of the calibration curve and then normalized by the corresponding measurement time (see Appendix). Table~\ref{Table_metrics_square} shows the sensitivity at the optimal point, corresponding to the largest slope of $F_{max}$, for different values of $R=\{15,\;20\}\,\mu\mathrm{m}$, while Fig.~\ref{fig:sensitivity} shows the sensitivity across the operating region of the sensor and how the latter changes with $R$. 
We stress that these values represent the sensitivity of an individual sensing cell and that for spatially varying fields the array provides $N_\text{cell}$ samples in parallel. The present proposal therefore targets field mapping with high spatial resolution rather than enhanced atom-number scaling beyond the standard quantum limit. We note that our choice of $R$ is constrained by the regions where $F_{max}$ changes substantially with the electric field, which is ultimately determined by the F\"orster resonance. Nonetheless, this freedom allows one to tune the trade-off between sensitivity and operational range depending on the size of the square cell, with larger $R$ giving higher sensitivity at the expense of a reduced operational range, whereas smaller $R$ provides a larger operational range but lower sensitivity (see Table~\ref{Table_metrics_square} and Fig.~\ref{fig:sensitivity}). Finally, we also point out that, due to the flexibility of different experimental setups~\cite{schlosser2023talbot}, one can obtain simultaneous operational ranges by creating different cell sizes either in the same array or in different planes, with the caveat that they operate at different sensitivities.
\begin{figure}[h!]
    \centering
    \includegraphics[width=\linewidth]{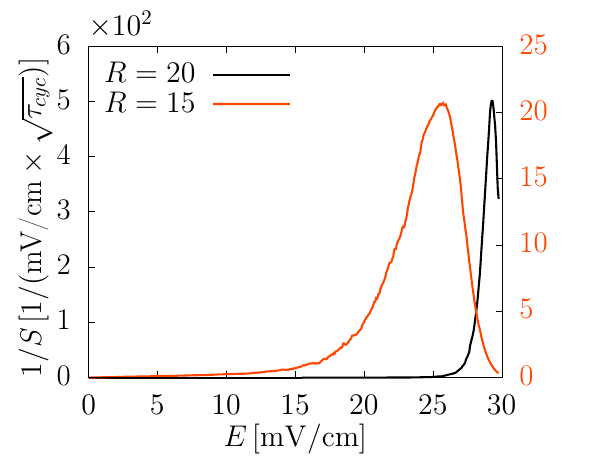}
    \caption{The reciprocal of the sensitivity $S$ as a function of the electric field $E$ for a square-cell configurations with interatomic separations $R = 20$ $\mu$m and $R = 15$ $\mu$m, corresponding to the left-hand and right-hand y-axes, respectively.}
    \label{fig:sensitivity}
\end{figure}

\gb{From a noise-reduction perspective, offsets and maximum amplitudes can also be calibrated using region of parameters where $F_{max}$ is constant for all electric fields around 0 or 1 for 
$R$ smaller and 
$R$ larger than the blockade radius, respectively.} These reference baselines can be used for normalization, while intermediate separations, comparable to the Rydberg blockade radii over the electric field range, serve as the primary probes for electric field sensing \cite{baselines}. This is important for distinguishing genuine field-dependent responses from background fluctuations or systematic offsets produced by spurious external fields such as magnetic gradients.

An alternative observable that provides spatial information about the electric field in the array at the micrometre scale is the density-density correlation (see Fig.~\ref{fig:density_density_N_19}). In this case, we just consider a linear array in which the electric field changes along the line of the atoms. In regions where the field is close to the Förster resonance, the correlator $\langle n_{i}n_{j}\rangle$ takes low values, consistent with a blockade radius exceeding the interatomic separation and suppressing simultaneous excitations of neighbouring atoms. In contrast, far from resonance, $\langle n_{i}n_{j}\rangle$ is large, indicating a reduced blockade radius that permits multiple excitations. By mapping these spatial variations in the correlator, the proposed setup can also resolve and characterize local electric field distributions.
}

\textit{Conclusion --} 
%
%
{\color{black} In this work, we demonstrated that an electric field gradiometer can be realized by measuring local Rydberg excitations in a 2D array, forming a sensor with micrometer-scale spatial resolution. The electric field locally modifies the Rydberg blockade radius via a F\"orster resonance, which enhances the electric-field dependence of the dipole-dipole interaction coefficient $C_3$. \la{We remark that F\"orster resonance is very important for the sensitivity of our protocols}: near resonance, small variations in the electric field produce pronounced changes in the interaction strength, thereby producing a steeper field dependence of the measured excitation signal.
%
%

}

Leveraging this phenomenon, we demonstrate the operating principles of an electric-field sensor based on a two-dimensional Rydberg network, as illustrated in Fig.~\ref{fig.tweezer_arrangement}. The sensing mechanism relies on monitoring the maximum fidelity $F_{\mathrm{max}}$ from~\eqref{eq.f_max}, which corresponds to the probability that all atoms in a given cell are excited to a Rydberg state 
For a fixed distance, the suppression of (or lack of thereof) $F_{\mathrm{max}}$ provides a direct signature of the local electric field, owing to its strong dependence of the Rydberg blockade radius. \blue{The tweezer array is arranged as a collection of independent square cells, each containing $N=4$ atoms separated by a distance $R$, while the separation between neighboring cells, $R_{\text{cell}}$, is chosen to be much larger than the intra-cell spacing. In this configuration, Rydberg interactions are effectively restricted to atoms within the same cell, while interactions between different cells remain negligible. By engineering cells with various atom-atom separations $R$, the array can also enable multiple, simultaneous measurements within a single experimental realization. Cells with separations close to the blockade radius exhibit maximal sensitivity to variations in the electric field, whereas cells with smaller separations provide wider operational range.}


Complementarily, the density–density correlator $\langle n_{i}n_{j}\rangle$ also provide information about the electric field spatial profile. In particular, low values of $\langle n_{i}n_{j}\rangle$ indicate strong suppression of neighbouring excitations, pointing to an effective interaction enhanced by the presence of a near-resonant electric field. 

Together, $F_{max}$ and $\langle n_{i}n_{j}\rangle$ serve as indicators of gradients of electric fields.
While the present implementation is capable of detecting spatially varying electric fields, its operational range is inherently constrained by the choice of atomic species and the specific Rydberg levels employed. These restrictions arise from the dependence of the blockade radius and level shifts on the principal quantum number and atomic properties. 
{\color{black}
{Further expansion of the operating range and improved sensitivity of the atomic network could be achieved by incorporating multiple atomic species~\cite{otto2020strong} or by exploiting different Rydberg states within the same array. 
}}

\gb{In a future application of this work, local electric fields might be used to manipulate Rydberg interactions and the Rydberg blockade.} This can be utilised in quantum simulation of models where certain atoms have stronger or weaker interactions, and can be achieved through local addressing of each atom with a designated laser beam.

\let\oldaddcontentsline\addcontentsline

\renewcommand{\addcontentsline}[3]{}

\medskip

\textit{Acknowledgements --} We thank Enrico C. Domanti for useful discussions. PK and LA acknowledge the Julian Schwinger Foundation grant JSF-18-12-0011. GB acknowledges financial support by the Federal Ministry of Research, Technology, and Space (BMFTR) [Grants 13N17366 and 13N17521], by the Deutsche Forschungsgemeinschaft (DFG -- German Research Foundation) [Grants BI 647/6-1, BI 647/6-2, and Priority Program SPP 1929 (GiRyd)], and the Honda Research Institute Europe. 


\bibliographystyle{iopart-num.bst}
\bibliography{lib.bib}

\newpage

\onecolumngrid
\section*{Appendix: Sensing electric fields through Rydberg atom networks}

\subsection{Rydberg Interactions, Rydberg Blockade and Förster Resonance}
\label{subsec:RydbergInteractionsRydbergBlockadeandFörsterResonance}

\begin{figure}[h]
    \centering
    \includegraphics[width=0.55\linewidth]{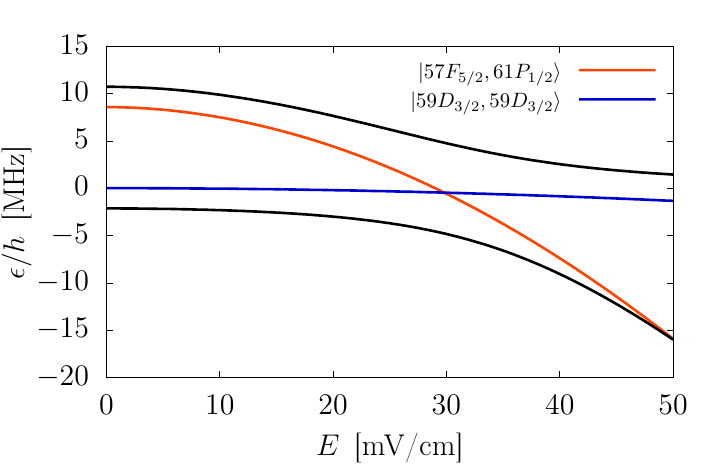}
    \caption{The relative energies $\epsilon$ in units of the Planck constant $h$ for the pair states $\ket{59D_{3/2},59D_{3/2}}$ and $\ket{57F_{5/2},61P_{1/2}}$
    when coupled (black) and uncoupled (coloured). The two atoms are separated by a distance $R = 8.1\mu$m and magnetic field $B_{z} = 0$ G.}
    \label{fig:Energy_Levels_in_E_Field}
\end{figure}

\noindent To provide a comprehensive analysis of the Förster resonance, we employ the ARC (Alkali Rydberg Calculator) toolbox to compute relevant data for the Rydberg levels of $^{87}$Rb. Specifically, we consider the pair-states $\ket{59D_{3/2}, \, 59D_{3/2}}$ and $\ket{57F_{5/2}, \, 61P_{1/2}}$ used in the main text, and evaluate their relative energies $\epsilon$ as a function of an applied electric field in the absence of a magnetic field ($B_z = 0 \,$ G). The resulting energy difference between the two pair states, referred to as the Förster defect $\delta(E)$, varies with the electric field and reaches zero at a particular value, marking the Förster resonance. As shown in Fig.~\ref{fig:Energy_Levels_in_E_Field}, this occurs at an electric field of approximately $E_{\text{res}} \approx 29.8,\text{mV/cm}$.
Additionally, the dipole matrix elements $d_{i}$ are calculated under each electric field condition, providing the dipole-dipole interaction as

\begin{equation}
    C_{3}(E) = \frac{d_{1}d_{2}}{4\pi\epsilon_{0}},
\end{equation}

\noindent where $\epsilon_{0}$ is the permittivity of free space. By using both $C_{3}(E)$ and $\delta(E)$ in conjunction with~\eqref{eq:eigenvalues}, the energy of the coupled system can be obtained, where an avoided crossing is observed at the Förster resonance. \\

\begin{figure}[h]
    \centering
    \includegraphics[width=0.75\linewidth]{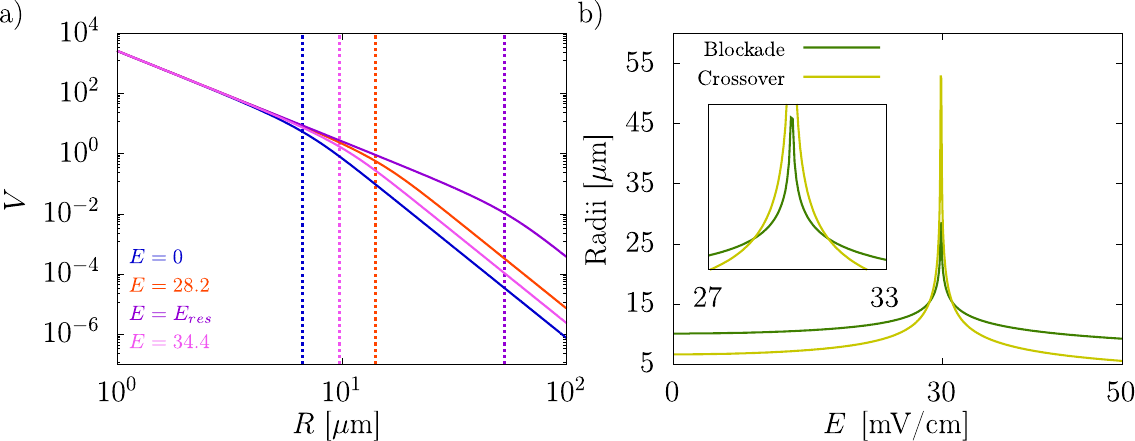}
    \caption{\idg{a} Effective interaction $V$ between two Rydberg atoms separated by a distance $R$ when subjected to electric field $E = 0$ mV/cm ($R_{c} = 6.68 \, \mu$m), $E = 28.2$ mV/cm  ($R_{c} = 14.15 \, \mu$m), $E = 29$ mV/cm  ($R_{c} = 52.95 \, \mu$m) and $E = 34.4$ mV/cm  ($R_{c} = 9.76 \, \mu$m). The change of behaviour between the dipole-dipole and Van der Waals interaction is recorded as $R_{c}$. \idg{b} The blockade radius (as defined in~\eqref{eq:Blockade}) and the interaction crossover radius, $R_{c}$ for different electric fields. The insert zooms into the region between $E = 27 - 33\, $mV/cm.}
    \label{fig:blockade_crossover}
\end{figure}

\noindent Figure~\ref{fig:blockade_crossover}\textbf{(a)} illustrates how the effective interaction $V$ varies with the separation between atoms. At short distances, the interaction is strong ($C_{3} > \delta$), resulting in a dipole-dipole–like behaviour with $V \propto R^{-3}$. As the separation increases and the interaction strength becomes weaker than the energy defect ($C_{3} < \delta$), the interaction crosses over to a van der Waals regime, scaling as $R^{-6}$. The point at which this transition occurs is identified as the crossover radius, 

\begin{equation}
    R_{c} = \left(\frac{C_{3}}{\hbar|\delta|}\right)^{1/3},
    \label{eq:interaction_crossover}
\end{equation}

\noindent which depends on the applied electric field. The largest value of the crossover radius is observed at the Förster resonance, $E = E_{\text{res}}$. This trend is further highlighted in Fig.~\ref{fig:blockade_crossover}\textbf{(b)}, where the crossover radius is plotted as a function of the electric field, showing a pronounced peak at the Förster resonance. For comparison, we also overlay the blockade radius, defined as~\cite{wu2023rydberg}

\begin{equation}
    R_{b} = \left(\frac{C_{6}}{\Omega}\right)^{1/6} = \left(\frac{C_{3}^{2}}{\hbar|\delta| \, \Omega}\right)^{1/6},
    \label{eq:Blockade}
\end{equation} 


\noindent where the angular component of the interaction is set to 1. Since the blockade radius depends on the energy defect $\delta$, it varies with the electric field. As the field approaches the Förster resonance ($E = E_{\text{res}}$), $\delta \to 0$ and the blockade radius reaches its maximum value. The nature of the interaction governing the blockade radius can be inferred by comparing it to the crossover radius. When the blockade radius is larger than the crossover radius ($R_{b} > R_{c}$), the interaction follows a van der Waals scaling, $V \propto R^{-6}$. Conversely, when the blockade radius is smaller than the crossover radius ($R_{b} < R_{c}$), the interaction is in the resonant regime and scales as $V \propto R^{-3}$—a situation that occurs predominantly near the Förster resonance.

\subsection{Maximum Fidelity for multiple atoms in a linear configuration}

\begin{figure}
    \centering
    \includegraphics[width=0.75\linewidth]{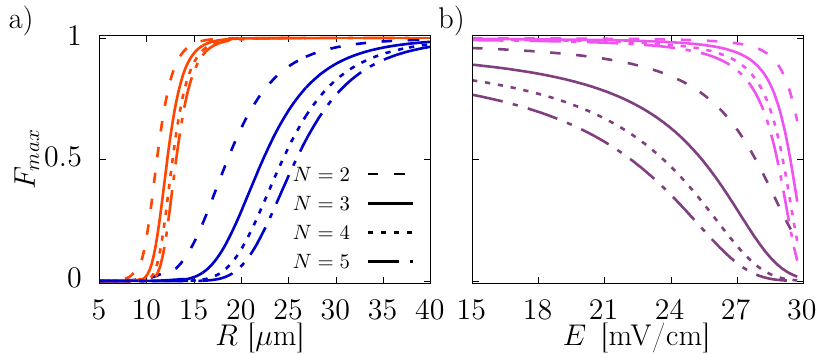}
    \caption{The maximum projection $F_{max}$ of $\ket{\Psi}$ to the Fock state $\ket{1_{1}, \cdots, 1_{N}}$ over the period $t \in (0, \tau = 2\pi/\Omega)$ for $N$ atoms, where $\Omega$ is the Rabi frequency. \idg{a} All atoms are subject to the same electric field (either $E = 0$ or $E = E_{res}$) and have atom separation $R$. \idg{b} All atoms have equal spacing (either $R = 15 \, \mu$m or $R = 20 \, \mu$m) with a uniform applied electric field.}
    \label{fig.f_max_many_atoms}
\end{figure}


\noindent The collective effect of the Rydberg blockade is examined by computing $F_{\text{max}}$ for systems with an increasing number of atoms. This serves as an indicator of how the blockade radius evolves with system size. Two electric field values are considered: $E = 0$ and $E = E_{\text{res}}$, with atoms arranged at a fixed separation $R$, as illustrated in Fig.~\ref{fig.f_max_many_atoms}\textbf{(a)}. As more atoms are introduced, $F_{\text{max}}$ shifts due to the additional interactions in the system. These collective contributions effectively enhance the interaction strength, resulting in an increased blockade radius. The extent of the shift varies with the electric field, as it determines the interaction scaling. At $E = 0$, interactions follow a van der Waals scaling ($V \propto R^{-6}$), whereas at the Förster resonance ($E = E_{\text{res}}$), they follow a resonant dipole-dipole scaling ($V \propto R^{-3}$). As a result, the stronger interactions at the resonance lead to a more pronounced shift in $F_{\text{max}}$ when atoms are added. However, this effect saturates gradually, as atoms farther away contribute less to the total interaction due to the distance dependence.\\


\noindent Furthermore, for fixed atom separations of $R = 15,\mu$m and $R = 20,\mu$m, the increase in blockade radius with the number of contributing atoms is illustrated in Fig.~\ref{fig.f_max_many_atoms}\textbf{(b)}. At a given electric field, $F_{\text{max}}$ decreases as the number of atoms grows. Examining the behaviour of the interaction $V$ at these separations, we observe that for small electric fields, the interaction scales as $R^{-6}$; however, as the electric field increases, it changes to a $R^{-3}$ scaling. This crossover occurs at a lower electric field for $R = 15,\mu$m compared to $R = 20,\mu$m, leading to a stronger effect on the blockade radius and thus a more pronounced shift in $F_{\text{max}}$ at the smaller separation.

\begin{figure*}[ht!]
    \centering
    \includegraphics[width=\linewidth]{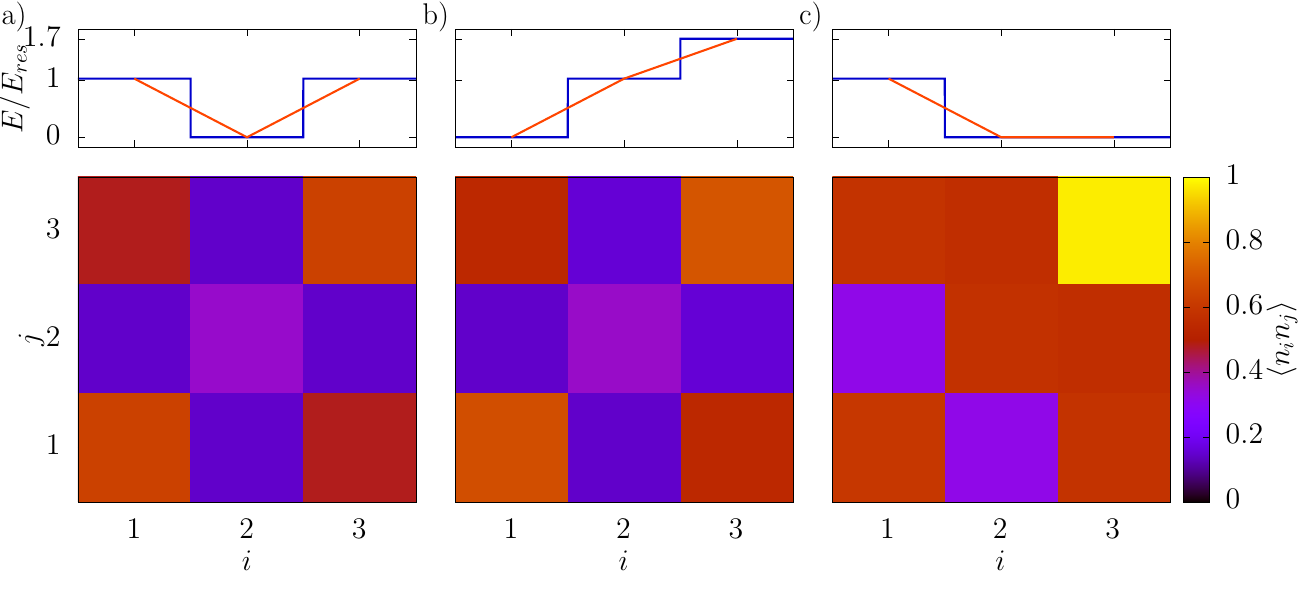}
    \caption{\textbf{Top:} The applied electric field configuration $E/E_{res}$ across $N = 3$ atoms, which are separated by a distance $R = 15 \, \mu$m. \textbf{Bottom:} The corresponding density-density correlator $\langle n_{i}n_{j}\rangle$ for the above electric field configurations recorded at maximum population of the fully excited state. \idg{a} $E/E_{res} = 1, 0, 1\,$. \idg{b} $E/E_{res} = 0, 1, 1.7\,$ (where $E/E_{res} = 1.7$ corresponds to the maximum electric field we consider $E = 50$ mV/cm). \idg{c} $E/E_{res} = 1, 0, 0$.}
    \label{fig:density_density_N_3}
\end{figure*} 

\subsection{Density-Density Correlator}
\label{sec:density_density_N_3}

\noindent The density-density correlator $\langle n_i n_j \rangle$ is calculated for a system of $N = 3$ atoms subject to a spatially varying electric field. Results for different electric field configurations are shown in Fig.~\ref{fig:density_density_N_3}. The diagonal elements represent the excitation probability at each site, while the super- and sub-diagonal elements reflect correlations between nearest neighbours. In configurations where one atom is near the Förster resonance, the correlator with its neighbour is suppressed due to the Rydberg blockade—i.e., the blockade radius exceeds the interatomic separation (as observed in Fig.~\ref{fig:density_density_N_3}\textbf{(a,b)}). Conversely, when neighbouring atoms are far from resonance, as in Fig.~\ref{fig:density_density_N_3}\textbf{(c)}, the correlators $\langle n_2 n_3 \rangle$ and $\langle n_3 n_2 \rangle$ are enhanced, indicating that the blockade is weaker than the separation and simultaneous Rydberg excitation is allowed.

{\color{black}


\subsection{Maximum Fidelity for $N = 3$ atoms in a linear formation}

\begin{figure}[h!]
    \centering
    \includegraphics[width=0.65\linewidth]{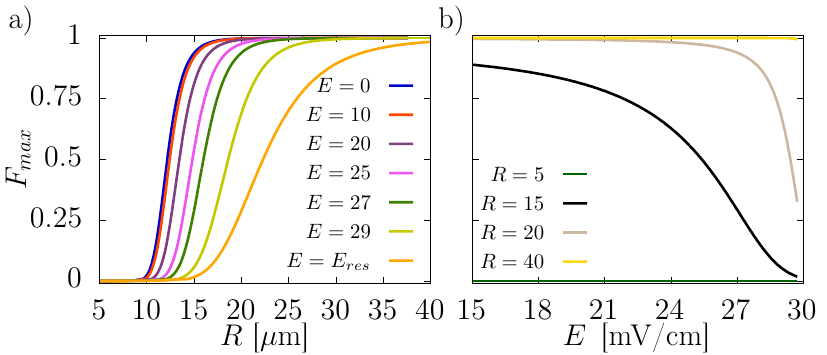}
    \caption{\idg{a} The maximum fidelity for populating the fully excited state, $F_{max}$ (~\eqref{eq.f_max}) for $N = 3$ over the time period $t \in (0, \, 2\pi/\Omega)$. The atoms have a fixed separation which increases from $R = 5 \, \mu$m to $R = 40 \, \mu$m for the different electric fields $E = (0, \, 10, \, 20, \, 25, \, 27, \, 29, \, E_{res})$ mV/cm.
    \idg{b} $F_{max}$ as a function of a uniform electric field across all $N = 3$ atoms for fixed separations $R = (5, \, 15, \, 20, \, 40) \, \mu$m.
    }
    \label{fig.f_max}
\end{figure} 

\noindent $F_{max}$ with the $N = 3$ atoms arranged in a linear configuration - see Fig.~\ref{fig.f_max}. Similar to the case of $N = 4$ atoms in a square configuration (presented in the main text), Fig.~\ref{fig.f_max}\textbf{(a)} shows that for weaker electric fields, the blockade radius is smaller, which allows all the atoms to be excited simultaneously. Analogously to the case shown in main text, Fig.~\ref{fig.f_max}\textbf{(b)} shows the $F_{max}$ for atoms separated by a distance $R$. For $R = 15 \mu$m, we observe a smaller operational range, and at $R = 20 \mu$m, the arrangement is less sensitive compared with the square configuration.

}

\color{black}

\subsection{Bayesian Statistics and Sensitivity}

\noindent Bayesian statistics is used to estimate an unknown parameter $A$ of a system which has been measured with the following outcome, $B = \{x_{1}(A), x_{2}(A), \cdots, x_{n}(A)\}$. Assuming the parameter exists within the space $A = [A_{min}, A_{max}]$, the use of Bayes' theorem estimates the Posterior distribution $P(A|B)$ around the target parameter. Bayes' Theorem is presented as, 
\begin{equation}
    P(A|B) = \frac{P(B|A)P(A)}{P(B)},
    \label{eq.Bayes_theorem}
\end{equation}
\noindent where $P(A)$ is the pre-known knowledge (distribution), $P(B|A) = p_{i}(x_{i}(A))$ is the conditional probability of making the observation $x_{i}(A)$. The posterior is an updated distribution, which is normalised via $P(B)$, and can be reused as $P(A)$ in another iteration of~\eqref{eq.Bayes_theorem}. \\

\subsubsection*{Parameter Estimation}

\begin{figure*}
    \centering
    \includegraphics[width=\linewidth]{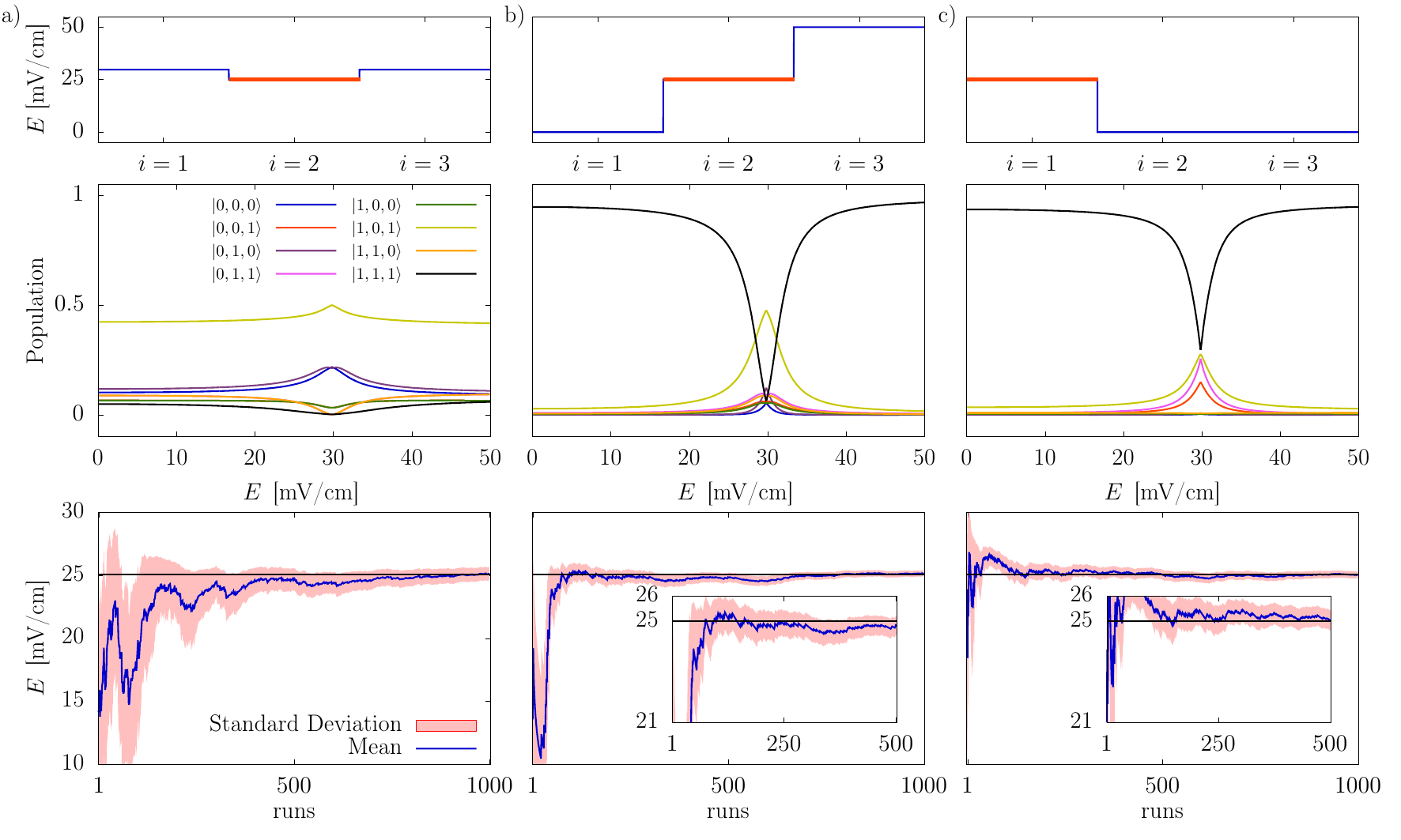}
    \caption{Bayesian analysis for $N = 3$ atoms separated by $R = 15 \, \mu$m to estimate a single value of an electric field. \textbf{Top:} The applied electric field configuration across the atoms, with the highlighted field corresponding to the parameter to estimate $E = 25$ mV/cm. \textbf{Middle:} The population of each of the state when the highlighted electric field is varied from $E = 0$ mV/cm to $E = 50$ mV/cm, recorded at $t = \pi/\Omega$. \textbf{Bottom}: The mean value posterior calculated through~\eqref{eq.Bayes_theorem} for $n_\mathrm{mes} = 1000$. The shaded area is the standard deviation of the posterior.}
    \label{fig.bayesian_statistics}
\end{figure*} 

\noindent We employ Bayes' theorem to estimate a single value parameter value. We use an array of $N = 3$ atoms, which all experience an independent electric field, considering three variations of configurations which are shown in the top row of Fig.~\ref{fig.bayesian_statistics}. For ease, we select the value to estimate as $E_{true} = 25$ mV/cm, which in the above description is $A$. However, as we are considering a system which is primarily unmeasured, therefore we assume no initial knowledge of $E_{true}$, the prior distribution is set to a uniform distribution $P(E_{i})$ where $E_{i} = [0, 50]$ mV/cm. To obtain the conditional probability, we investigate the state populations (amplitudes) when the electric field on the atom we want to measure varies over $E_{i}$ - this can be seen in the middle row of Fig.~\ref{fig.bayesian_statistics}. In doing this, we denote $x_{i}$ as each of the states (which for $N = 3$ are eight in total) and their population correspond to the probability $p_{i}$. \\

\noindent Applying Bayes' theorem for the three electric field configurations in Fig.~\ref{fig.bayesian_statistics}, we iterate through the algorithm 1000 runs to show the estimation converges on $E_{true}$. If the applied electric field on either of the two auxiliary atoms is close enough to the resonance, so that the resulting blockade radius is greater than the atom separation, the population of the fully excited state remains low and relatively constant for all $E$. This situation is shown in Fig.~\ref{fig.bayesian_statistics}\textbf{(a)}, where the difficulty arises that due to the non-varying state populations, Bayes' theorem cannot converge on $E_{true}$ as the probability of each measurement is similar for all values of $E$. However, suppose the auxiliary atoms are subject to an electric field where the blockade radius is smaller than the atom separation. In that case, we observe a change in the population of the fully excited state as $E$ changes, because this atom now dominates whether a blockade occurs or not. This situation is given in both Fig.~\ref{fig.bayesian_statistics}\textbf{(b, c)}, where, due to the changing population for the applied $E$, Bayes' theorem can estimate $E_{true}$ for a small number of runs. \\

\subsubsection*{Calculation of the Sensitivity}

\noindent We utilise Bayesian statistics to estimate the values of $F_{max}$ in Fig.~\ref{fig_f_max_N_4}\textbf{(b)} for both $R = 15 \mu$m and $R = 20 \mu$m. Then, for each electric field, we iterate through~\eqref{eq.Bayes_theorem} for $n_\mathrm{mes} = 120$ runs, obtaining a mean value of $F_{max}$ along with an associated error, $\sigma_{error}$, in the measurement. From this, the sensitivity of recording each electric field can be calculated through:
\begin{equation}
    S = \frac{\sigma_{error}}{\abs{\frac{d F_{max}}{d E}}} \times n_\mathrm{mes} \times \sqrt{\tau_{cyc}}
\end{equation}
where $\tau_{cyc}$ is the experimental cycle time. \gb{Estimates for typical cycle time are given in \cite{schaeffner2024Bfieldsensor}.}

\subsubsection*{Calculation of the Operational Range}

\noindent The operational range quantifies the span of electric-field values over which the sensor provides a useful response. To estimate it, we use the curves of the reciprocal sensitivity, $1/S(E)$, shown in Fig.~\ref{fig:sensitivity}, and treat them as truncated, off-centred distributions over the electric field after normalization. Specifically, we define a normalized weighting function
\begin{equation}
p(E) = \frac{1/S(E)}{\int dE\, [1/S(E)]}.
\end{equation}
Using this distribution, we compute the mean electric field
\begin{equation}
\mu = \int dE\, E\, p(E),
\end{equation}
and the variance
\begin{equation}
\sigma^2 = \int dE\, (E - \mu)^2\, p(E) .
\end{equation}
The operational range is then defined as $2\sigma$, which provides a characteristic width of the region over which the sensor remains responsive. Applying this procedure to the two datasets in Fig.~\ref{fig:sensitivity} yields the operational ranges reported in Table \ref{Table_metrics_square} for atom separations $R = 15\,\mu\mathrm{m}$ and $R = 20\,\mu\mathrm{m}$.

\end{document}